\documentclass[%
 aip,
 rsi,%
reprint,%
 graphicx,
 amsmath,amssymb,
]{revtex4-1}

\usepackage{siunitx}% Include figure files
\usepackage{graphicx}% Include figure files
\usepackage{dcolumn}% Align table columns on decimal point
\usepackage{bm}% bold math
\usepackage[english]{babel}
\makeatletter
\adddialect\l@ENGLISH\l@english
\adddialect\l@en\l@english
\makeatother
\usepackage{tabularx}
\usepackage{hyperref}
\usepackage[table]{xcolor}
\usepackage{array, ragged2e}
\hypersetup{
    colorlinks,
    linkcolor={red!50!black},
    citecolor={blue!50!black},
    urlcolor={blue!80!black}
}
\begin{document}

\preprint{AIP/123-QED}

\title[Lithium vapor box similarity experiment]{Design and measurement methods for a lithium vapor box similarity experiment}% Force line breaks with \\
%\title[Lithium vapor box similarity experiment]{Vapor flow measurements on a Lithium Vapor Box Similarity Experiment}% Force line breaks with \\
\thanks{Contributed paper published as part of the Proceedings of the 22nd Topical Conference on High-Temperature Plasma Diagnostics, San Diego, California, April, 2018.\\}

\author{J. A. Schwartz}
\email{jschwart@pppl.gov}
\author{E. D. Emdee}
\affiliation{Princeton University, Princeton, New Jersey 08540, USA}
\author{M. A. Jaworski}%
\affiliation{Princeton Plasma Physics Laboratory, Princeton, New Jersey 08543-0451, USA}
\author{R. J. Goldston}%
\affiliation{Princeton University, Princeton, New Jersey 08540, USA}

\date{\today}% It is always \today, today,

\begin{abstract}
The lithium vapor box divertor is a concept for handling the extreme divertor heat fluxes in magnetic fusion devices. In a baffled slot divertor, plasma interacts with a dense cloud of Li vapor which radiates and cools the plasma, leading to recombination and detachment. Before testing on a tokamak the concept should be validated: we plan to study detachment and heat redistribution by a Li vapor cloud in laboratory experiments. Mass changes and temperatures are measured to validate a Direct Simulation Monte Carlo model of neutral Li. The initial experiment involves a \SI{5}{cm} diameter steel box containing \SI{10}{g} of Li held at 650$^\circ$C as vapor flows out a wide nozzle into a similarly-sized box at a lower temperature. Diagnosis is made challenging by the required material compatibility with lithium vapor. Vapor pressure a steep function of temperature, so to validate mass flow models to within 10\%, absolute temperature to within \SI{4.5}{\kelvin} is required. The apparatus is designed to be used with an analytical balance to determine mass transport. Details of the apparatus and methods of temperature and mass flow measurements are presented.
\end{abstract}

 %Use showkeys class option if keyword
\maketitle

\section{\label{sec:intro}Introduction}

The divertor heat flux in future toriodal fusion reactors will likely require a detached plasma\cite{maingi_fusion_2015}. 
While experiments on present tokamaks have demonstrated detachment via gas puffs, \cite{leonard_plasma_2018} gas may build up in the divertor and cause the detachment front to move upstream, ending in a MARFE and/or disruption \cite{hutchinson_thermal_1994,lipschultz_sensitivity_2016}. A method of producing stable detachment is required, such as the lithium vapor box (LVB) divertor \cite{goldston_lithium_2016, goldston_recent_2017}.
%, aims to produce stable detachment by means of a localized impurity cloud inside a baffled slot.
%Inside the bottom, hottest chamber of the vapor box, lithium evaporates from a liquid surface at \SIrange{815}{1037}{\kelvin} and enters the plasma, causing it to cool via radiation and ultimately recombine.
%The additional baffles and chambers, at lower temperatures, act to condense the vapor and prevent a large neutral efflux to the main plasma chamber.

%\begin{figure}
%\includegraphics[width=0.5 \columnwidth]{goldstonVBD}% Here is how to import EPS art
%\caption{\label{fig:goldstonVBD} Poloidal cross section of a lithium vapor box divertor. The divertor leg plasma (magenta) terminates in bottom of the slot where lithium vapor is most dense. The additional baffles and boxes constrain lithium vapor efflux to the main plasma.}
%\end{figure}
%
Before testing on a toroidal device, properties and behavior of the LVB can be tested on a smaller scale, such as a linear divertor test facility \cite{van_eck_operational_2014}.
Key parameters are the density of Li vapor necessary for detachment, 
the dispersal of the plasma heat flux during detachment,
the amount of Li carried upstream through the plasma,
and the amount of neutral Li efflux to the main chamber, some of which can be studied in linear plasma devices.

A preliminary experiment at PPPL has been designed to measure Li efflux without plasma present, as would be between plasma diacharges. This is done in order to validate a Direct Simulation Monte Carlo \cite{bird_monte_1978, bird_molecular_1994} (DSMC) code SPARTA\cite{gallis_direct_2014} for use with evaporating and condensing lithium systems in the transitional flow regime.

\begin{figure}
\includegraphics[width=0.70 \columnwidth]{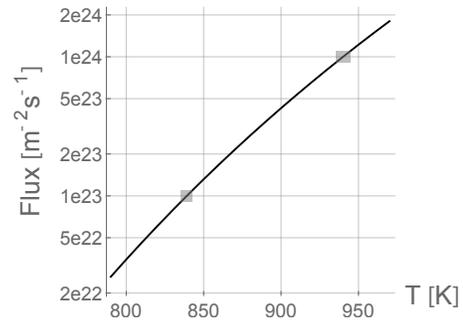}% Here is how to import EPS art
\caption{\label{fig:fluxplot} Equilibrium Li vapor flux $\Gamma$ (\autoref{eq:vaporFlux}) at temperatures relevant to the hottest box. Gray rectangles show the temperature range for a $\pm10\%$ range in $\Gamma$.}
\end{figure}

The experimental device, scaled with flow parameters relevant to those that would be found on a tokamak, is a linear string of three \SI{5}{cm}-diameter cylindrical boxes, each heated to different temperatures up to $> \SI{900}{\kelvin}$.
An initial quantity of Li, \SI{10}{\gram}, is placed in the hottest of the boxes, where it evaporates and condenses in the cooler boxes.
Lithium is reactive and corrosive to many materials, so it is difficult to directly monitor properties of the vapor flow inside the device.
%Instead, the boxes are weighed before and after a heating, temperature hold, and cooling cycle to determine the mass transferred.
Instead, the boxes are weighed before and after a high temperature cycle to determine the mass transferred.
Flow rates are deduced from the mass change $\Delta m$ over a duration $\Delta t$.
The boxes are designed to separate to be weighed.
Estimates of flow rates in Reference \onlinecite{goldston_lithium_2016} show reductions by an order of magnitude per box and typical flow rates of a few \si{\milli\gram\per\second}, so lightweight boxes and accurate mass change measurements are necessary.
Mass flow rates in the device are a strong function of temperature: as shown in \autoref{fig:fluxplot}, a \SI{4}{\kelvin} surface temperature change at \SI{873}{\kelvin} produces an increase in vapor flux of 10\%.
A 21-channel thermocouple (TC) system is used to measure box wall temperatures.

Details of the experimental design including lightweight, separable boxes and considerations for the design of the TC system and error estimates are discussed below.

\section{\label{sec:needed}Theory and requirements}
A simple model for divertor power dissipation\cite{goldston_recent_2017} assumes that a divertor leg surrounded by vapor experiences an inward flux of neutral atoms and a constant amount of cooling per atom, $\epsilon_c$.

Depending on power, geometry and $\epsilon_c$ (see Equation 2 of Reference \onlinecite{goldston_recent_2017}),
%\footnote{$P = 4 \pi R \, l_p \epsilon_c \, \Gamma$. Power $P=$ \SIrange{150}{300}{\mega\watt}, major radius $R=$\SIrange{5}{7}{\meter}, poloidal vapor box length $l_p = $\SIrange{0.5}{1}{m}, Cooling power $\epsilon = $\SIrange{10}{200}{\electronvolt} per atom.},
Li vapor number fluxes to extinguish a DEMO divertor plasma are estimated to be of magnitude \SIrange{5e22}{6e24}{m^{-2}s^{-1}}, corresponding to one-way fluxes (defined below) of equilibrium vapor with temperatures of \SIrange{800}{1100}{K} respectively.

The one-way flux $\Gamma$ through a surface of an equilibrium ideal gas is given by 
\begin{equation}
  \label{eq:vaporFlux}
  \Gamma = \frac{1}{4}n\bar{v} = \frac{p}{ \sqrt{2 \pi m k_b T}}
\end{equation}
with $\Gamma$ in \si{m^{-2}s^{-1}}, density $n$ in \si{m^{-3}}, average velocity $\bar{v}$ in \si{m/s}, pressure $p$ in \si{Pa}, molecular mass $m$ in \si{\kilo\gram}, $T$ in \si{\kelvin}, and Boltzmann constant $k_b$.

%Li vapor pressure $p_\text{Li}(T)$ is nearly an exponential function of temperature\cite{davison_complication_1968,hicks_evaluation_1963,maucherat_pression_1939, browning_assessment_1985}  temperature change $\delta T$ corresponds to a larger relative change in $\Gamma$.
Li vapor pressure $p_\text{Li}(T)$ is a steep function of temperature\cite{davison_complication_1968,hicks_evaluation_1963,maucherat_pression_1939, browning_assessment_1985} and at temperatures relevant to this experiment, $\delta \Gamma / \Gamma > \delta T / T$.
For example, errors $\delta T$ for which $\delta \Gamma = 10\%$, $\Gamma(T_0 + \delta T)/\Gamma(T_0) = 1.1$, range from \SI{3.5}{K} at \SI{815}{K} and \SI{4.5}{K} at \SI{920}{K} to \SI{5.8}{K} at \SI{1037}{K}. 

%\begin{table}
%\caption{\label{tab:tabFluxErr}

%  Temperature increases $\delta T$ for which equilibrium vapor fluxes $\Gamma$ increase by a 10\%.
%}
%\begin{ruledtabular}
%\begin{tabular}{rddd}
%$T_0$ [K]      & 815 & 919 & 1037 \\
%$\delta T$ [K] & 3.5 & 4.5 & 5.8 
%\end{tabular}
%\end{ruledtabular}
%\end{table}
The vapor box operates in regimes of intermediate Knudsen number, $\text{Kn} = \lambda_\text{MFP}/L$ from 0.01 to 1.
In this transitional flow regime, DSMC models are more accurate than fluid models since they simulate particles with finite mean free paths.
A DSMC model for the vapor box using the code SPARTA\cite{gallis_direct_2014} is under development.
The typical scale of vapor pressures, flows through the device and evaporation fluxes are set by $p_\text{Li}(T)$ and $\Gamma$.

Therefore, validation of flow rates computed with DSMC to 10\% using an experiment with a first box at \SI{920}{\kelvin} will require measurements of mass flow through the boxes to better than 10\% as well as measurements of liquid Li surface temperatures to better than $\pm\SI{4.5}{\kelvin}$.
\section{\label{sec:apparatus}Apparatus and Approach}

\begin{figure}
\includegraphics[width=\columnwidth]{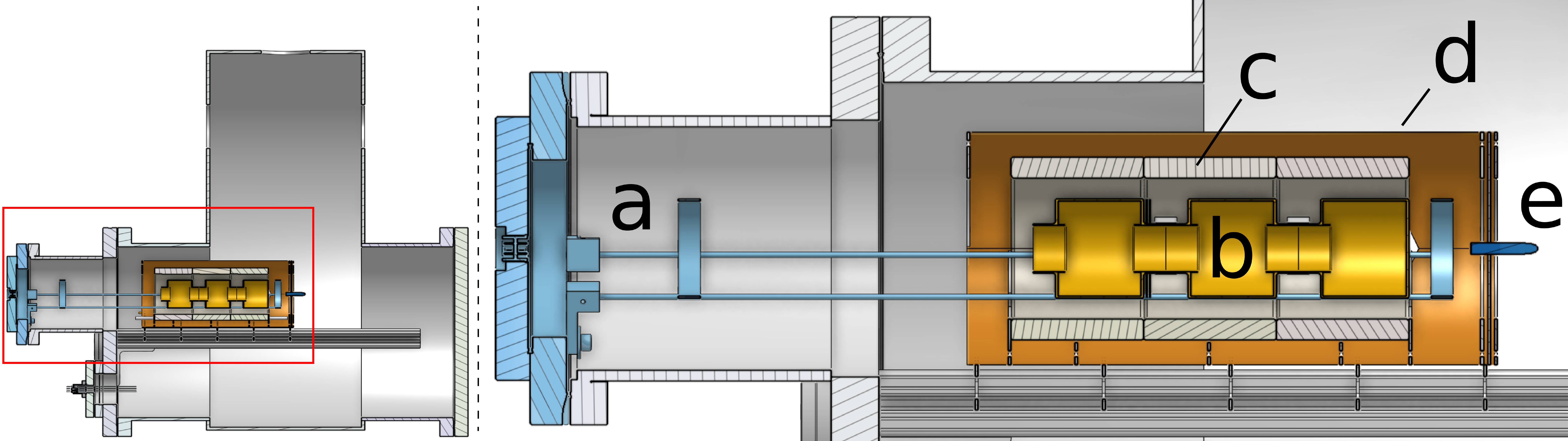}
\caption{\label{fig:transfer_assy} 
Left, a cross section of the apparatus. Right, detail of red rectangle.
(a) In blue, a flange attached to a 50-pin electrical feedthrough and to rods that support the boxes.
(b) In yellow, three vapor boxes.
The `box 1', at right, is hottest.
Vapor will flow from right to left.
Parts (a) and (b) are moved to and from the glovebox together.
(c) In white, the three ceramic radiant heaters.
(d) In orange, the SS shim stock `outer shield'.
(e) In dark blue, the gas feed tube (right end curved out of the plane).
}
\end{figure}

\autoref{fig:transfer_assy} shows the design of the apparatus, in the configuration with the `transfer assembly' bolted into the vacuum chamber.
The transfer assembly, (a) and (b) in the image, is composed of the Li-containing boxes, a 50-pin electrical feedthrough, and supports to hold up the boxes.
This high-pin-count feedthrough has pins made of a Be-Cu alloy.
During an experiment the transfer assembly is moved from an argon glovebox to the vacuum chamber and back.
First in the glovebox, \SI{10}{\gram} of solid Li is placed in `box 1' shown on the right of the image.
Each box is weighed to establish an initial mass.
The boxes are joined together and placed on the transfer assembly.
Then in the vacuum chamber, the boxes are heated by radiant heaters (c), held for a `temperature soak' period of typically 30 minutes while Li flows from box 1 to the other boxes and out, then cooled by a He gas feed from the tube (e).
Finally the transfer assembly is moved back to the glovebox, and the boxes are separated and reweighed to determine $\Delta m$.

As shown in \autoref{eq:masschange}, the total mass change is equal to the integral of the difference of flows in and out of the box over the heating-cooling cycle.
\begin{equation}\label{eq:masschange}
  \Delta m = \int (\dot{m}_\text{in} - \dot{m}_\text{out}) \; dt
\end{equation}
Typical experimental parameters are given in \autoref{tab:table1}. 
\begin{table}
\caption{\label{tab:table1}
Example steady state parameters for vapor box system, computing using the `choked flow model' \cite{goldston_lithium_2016} with flow reduced by 2 to match preliminary DSMC simulations.}
\begin{ruledtabular}
\begin{tabular}{lddd}
Box number &1&2&3\\
\hline
Flow Knudsen number & 0.0448 & 0.136 & 0.435 \\
Box temperature [K] & 919 & 849 & 803\\
Heater temperature [K] & 1071 & 774 & 778 \\
Box emissivity  & 0.3 & 0.85 & 0.85 \\
Heater emissivity  & 0.85 & 0.85 & 0.85 \\
Initial Li load [g] & 10 & 0 & 0 \\
Lithium efflux [mg/s] & 4.0 & 1.2 & 0.36 \\
Over 30 minutes [g] & 7.1 & 2.2 & 0.66 \\
Net $\Delta m$ [g] & -7.1 & +5.0 & +1.5 \\
Net latent heat flow [W] & -91 & +63 & +19 \\
\end{tabular}
\end{ruledtabular}
\end{table}
The justification for the design follows.

Liquid Li is a highly reactive substance, but is compatible with stainless steel 304 (SS), a common alloy.
The $\Delta m$ over time method allows all Li-facing surfaces to be made of SS, with no need for windows into the vapor-filled container.
The method also imposes three constraints: the weighed boxes should be as light as possible relative to $\Delta m$, the boxes must be separable in order to be weighed individually, and the boxes must be heated and cooled quickly to minimize flow during the transient periods.
In addition, the initial Li load (and therefore required soak time) should be minimized for safety.
\begin{figure}
\includegraphics[width=0.7\columnwidth]{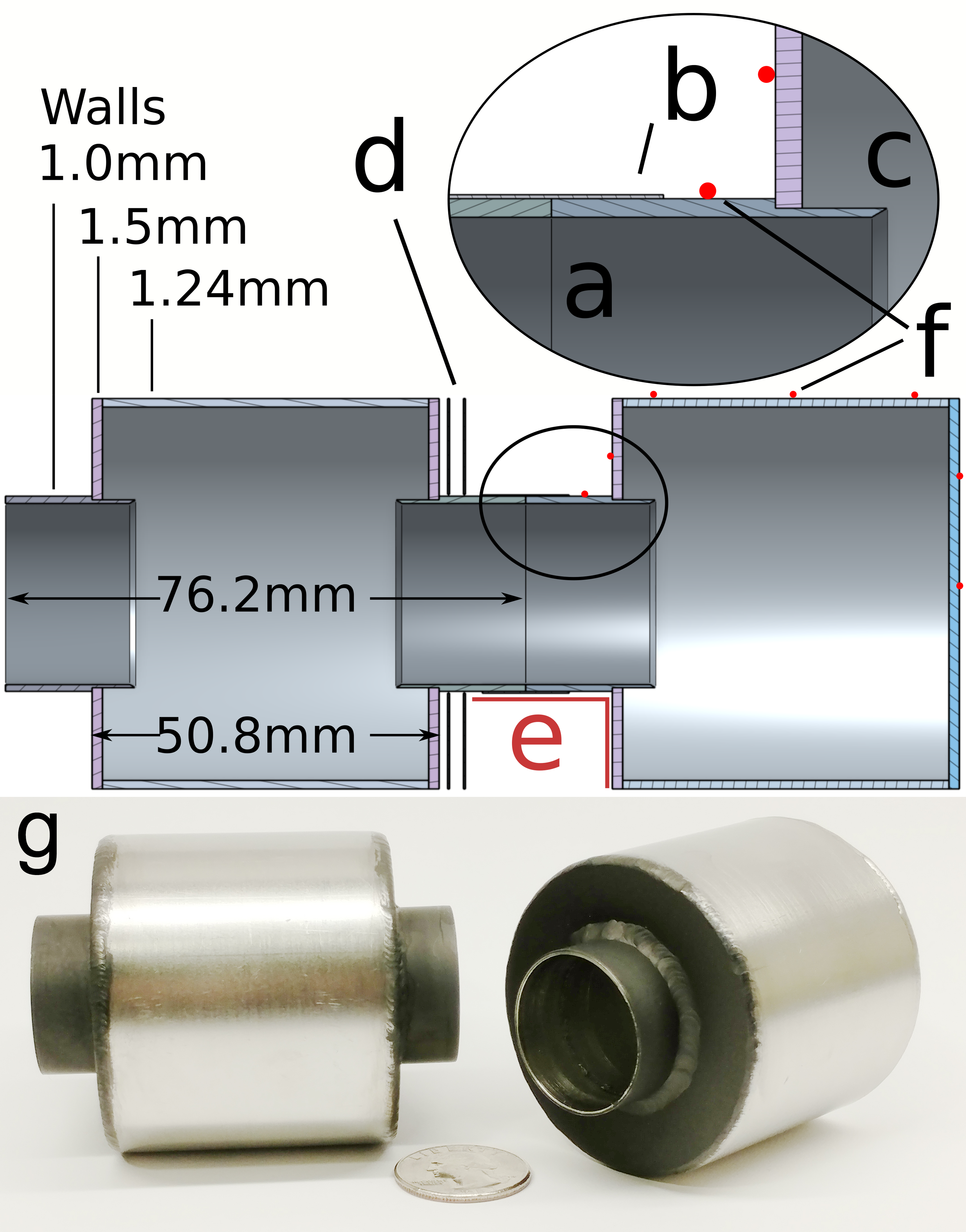}% Here is how to import EPS art
\caption{\label{fig:cut} Design features of the vapor boxes. (a) Butt joint between boxes, to allow separation. Other joints are welded. (b) \SI{0.05}{mm} SS shimstock wrapped around joint for friction. (c) Nozzle inner protrusion with knife edge, to prevent liquid Li wicking through the joint. (d) SS shim stock reflective shields. (e) Lines: high-emissivity coatings keep the joint region hot to prevent condensation. (f) Dots: example TC tip locations. (g) Photo with US quarter for scale.}
\end{figure}

\autoref{fig:cut} shows a detail of the box.
The box walls are 1 to \SI{1.5}{mm} thick  SS.
There is one `end box' and two `central boxes'.
The outer diameter of the main cylinder and nozzles are \SI{57.2}{mm}, and \SI{28.6}{mm}, respectively.
The total length of a central box is \SI{76.2}{mm}.

Between boxes are butt joints, wrapped in \SI{0.05}{mm} SS shimstock to prevent vapor leakage.
Joint regions have a high-emissivity coating (Aremco 840-M) painted on to keep them hotter than the previous box and prevent condensation.
To prevent \textit{liquid} Li wicking through the joint, a knife edge protrudes inward from the nozzle.

Each box is instrumented by up to 7 permanently-attached TCs, chosen for their small size, low wire count, and accuracy at high temperatures.
There are Macor connectors in the TC circuit between the boxes and feedthrough so the boxes can be weighed separately. 

Including the TCs and connectors, the end and central boxes weigh 160 and \SI{165}{g}, respectively.

The boxes are heated by thermal radiation to save weight: no bond to a heater is required. 
The heaters (Thermcraft, Inc, model RH251, similar to those on LITER\cite{kugel_lithium_2010}) have rated power \SI{720}{W} per cylindrical segment.
The inner faces are coated (Aremco 840-C) for high emissivity.
The box and heater lengths are equal so that temperatures of each box can be independently controlled.
In the space between the boxes and heaters, TC wires are routed and gas flows during cooling.
%The space between boxes and heaters is used to route TC wires and for the cooling gas feed to flow through. 

Outside the heaters is a cylindrical SS shim stock `outer shield', \autoref{fig:transfer_assy} (d), of diameter \SI{143}{mm} and thickness \SI{0.25}{mm}.
It is a physical support for the heaters, thermal radiation shield, and duct for the cooling gas.

A gas feed of He or Ar is supplied by a tube near the hottest box, \autoref{fig:transfer_assy} (e).
The system is designed for a He flow rate of 80 standard \si{L\per\minute} to cool down the boxes after the temperature soak phase.
An Ar purge prevents oxidation of Li surfaces during transfer assembly insertion and removal. 
%A \SI{500}{\liter\per\second} turbopump except during He flow cooling, when the \SI{780}{\liter\per\minute} roughing pump to maintain sub-atmospheric pressure.
The chamber is pumped by a \SI{780}{L\per\minute} roughing pump during He flow and by a \SI{500}{L\per\second} turbopump during the heating and soak phases.

\section{\label{sec:results}Measurement systems and errors}
%\begin{table}
%\caption{\label{tab:ttc} Components in the thermocouple system. NI is National Instruments. SMPW and SMTC are parts from Omega Engineering. STD and SLE are `Standard Limits of Error' and `Special Limits of Error' material. }
%\begin{ruledtabular}
%\begin{tabular}{llll}
%part & size & length & material \\
%\hline
%NI 9214 reader & & & \\
%Wire & 24 AWG &  \SI{15}{cm} & SLE \\
%SMPW connectors & & & STD \\
%Wire & 20 AWG & \SI{5}{\meter} & SLE \\
%SMTC connector sockets & & & STD \\
%Vacuum feedthrough pins & \SI{0.5}{mm} & \SI{5.}{\milli\meter} & Copper \\
%SMTC connector sockets & & & STD \\
%Wire & 20 AWG & \SI{300}{mm} & SLE \\
%SMTC connector pins & & & STD \\
%SMTC connector sockets & & & STD \\
%Wire & 34 AWG & \SI{20}{mm} & SLE \\
%Omegaclad XL probe & \SI{1}{mm} sheath & \SI{475}{mm} & SLE
%\end{tabular}
%\end{ruledtabular}
%\end{table}
Thermocouples (TCs) are used to measure temperatures of the boxes.
The system has 32 channels, up to 25 of which are measurements on the transfer assembly.
Type N TCs are used throughout the system for their stability compared to Type K\cite{burley_nicrosil_1982, anderson_decalibration_1982}.
`Special limits of error' (SLE) grade material\cite{noauthor_standard_nodate} is used when available. 
Inside the vacuum feedthrough, the wires are made into three bundles of 7 TCs each: one for each box.
Pairs of Macor 15-pin subminiature-D connectors allow the boxes to be weighed separately.
Finer wires at the readers and hot junction reduce thermal shunting, while thicker wires minimize circuit resistance.
The grounded probes are sheathed in `Omegaclad XL' for greater drift stability compared to SS or Inconel sheathed TCs.
The tips are spot welded to the boxes beneath a small cover of SS shimstock to decrease influence on measured temperature by thermal radiation.

An analytical balance, Mettler Toledo MS-603-TS, is used to weigh the boxes.
It has specified maximum weight of \SI{620}{g}, readability and repeatability of \SI{1}{\milli\gram} and minimum sample weight of \SI{140}{\milli\gram}.
%We have not yet tested it in the Ar glovebox. 

\subsection{Thermocouple Measurement Errors}
Each component of the system can cause errors.
The National Instruments 9214 reader has errors due to cold junction compensation (CJC) as well as noise, amplifier gain and offset.
Preliminary tests using a TC calibrator (Omega CL543B) show no deviations larger than \SI{0.2}{\kelvin}. Variation may occur between experiments, and can be tested by repeated calibrations.

In each circuit are several lengths of TC wire.
The SLE specification for Type N at \SI{923}{\kelvin} is $\pm 0.4\%$ of reading: $\pm\SI{2.6}{K}$.
This can be improved by calibration: we plan to purchase TC probes with NIST-traceable calibration to $\pm\SI{0.8}{\kelvin}$ at \SI{923}{\kelvin}.
Each lead wire segment also requires calibration. Four segments, each with two ends, plus the cold end of the probe, each calibrated to $\pm\SI{0.2}{K}$ sum in quadrature to $\pm\SI{0.6}{K}$.
Temperature differences across the Be-Cu alloy feedthrough pins can be subtracted by a circuit through a TC-material feedthrough\cite{ziemke_correction_2006}. The error in subtraction is estimated equal to the NI 9214 read error. The feedthrough temperature difference can be cross-checked by measurements of temperature on inner and outer feedthrough surfaces made by another method, such as resistance temperature detectors. 
%Any $\Delta T$ across the pins would be unaccounted by TCs. 
%Assuming\cite{lide_thermal_2008} the thermal conductivity is \SI{100}{\watt\per\meter\per\kelvin} a gradient of \SI{1}{K} over the 50 pins could be sustained by \SI{0.3}{\watt}, so a considerable $\Delta T$ is possible.
%This error can be mitigated using a TC circuit through both a low-channel count TC feedthrough and back through the BeCu pins \cite{ziemke_correction_2006}.

Drifts in base-metal TCs occur over hours to weeks, depending on construction, temperatures and environmental conditions, and generally cannot be determined by a recalibration\cite{noauthor_standard_nodate}.
Estimates for drifts due to reversible changes\cite{bentley_thermoelectric_1989}, thermal cycling\cite{belevtsev_stability_2003}, and irreversible changes\cite{bentley_irreversible_1989} are shown in \autoref{tab:tableError}.

Li evaporation will cool box wall inner surfaces relative to outer ones.
With the expected (\autoref{tab:table1}) \SI{90}{\watt} power flow, assumed through only the main cylindrical surface, and SS thermal conductivity\cite{sweet_thermal_1987} of \SI{25}{\watt\per\meter\per\kelvin}, the temperature difference is \SI{0.5}{K}, with estimated error \SI{0.2}{K}.

TC cable between a box and heater absorbs radiation, causing thermal shunting to the hot junction.
Absorbed heat and thermal gradients can be reduced by reflective shielding and by spot welding multiple locations along the TC to the box.
A test stand is being used to refine error estimates, listed as \SI{0.5}{K}.

Contributions to total uncertainty are summarized in \autoref{tab:tableError}.
The sum, \SI{4.6}{K}, is larger than the \SI{4.5}{K} required to model flow to 10\%, so better quantification of systematic errors and correlations, through modeling and experience with the test system, will be required.
\begin{table}
\caption{\label{tab:tableError}
  Estimated uncertainties in the thermocouple system. Uncertainties are treated as systematic with the exception of the first, marked $^R$ for random.
}
\begin{tabular}{m{53mm} m{25mm}}
\hline
\hline
\vspace{3pt}
  \centering  Source & \vspace{3pt}Uncertainty [K] \\

\end{tabular}
\begin{tabular}{lcl}
\hline
NI 9214 reader &\quad\quad &0.2$^{R}$ \\
Cancellation of $\Delta T$ across feedthrough pins&\quad\quad& 0.2 \\
Probe (NIST-traceable calibration at \SI{923}{K}) &\quad\quad& 0.8 \\
%Probe (calibration services at \SI{923}{K}) & 0.8 \\
Wires (quadrature of 9 calibrations to $\pm\SI{0.2}{K}$) &\quad\quad& 0.6 \\
Reversible changes at \SI{923}{K}, 100 hours \cite{bentley_thermoelectric_1989} &\quad\quad& 1 \\
After 30 cycles to \SI{1360}{\kelvin} \cite{belevtsev_stability_2003} &\quad\quad& 0.9 \\
Irreversible changes at \SI{923}{\kelvin}, 100 hours \cite{bentley_irreversible_1989} &\quad\quad& 0.2 \\
Box wall $\Delta T$ between outer and inner surfaces &\quad\quad& 0.2 \\
Estimated thermal shunting &\quad\quad& 0.5 \\
\hline
Sum &\quad\quad& 4.6 \\
\hline \hline
\end{tabular}
\end{table}
\subsection{Mass measurement errors}
During transfer between the vacuum chamber and glovebox the boxes will be filled with Ar.
The last box will be plugged with a stopper, but air leaks through joints may oxidize Li surfaces.
%Oxidation mass gain is estimated by two methods.
%The first is by arbitrarily assuming 100 Li monolayers (\SI{17}{\nano\meter}) would oxidize by the reaction $4\text{Li} + \text{O}_2 \to 2 \text{Li}_2 \text{O}$ for a mass gain of \SI{1.55}{\milli\gram}.
%The second is by estimating oxygen diffusion through a small leak, chosen arbitrarily to be a radius $b = \SI{1}{mm}$ hole. Using a diffusion coefficient of $\text{O}_2$ into Ar, $D= \SI{0.189}{\centi\meter\squared\per\second}$ the total leaked quantity as a function of time is
%\begin{equation}
%  V_\text{leak}(t) = 4 \pi b D t + 8 b^2 \sqrt{\pi D t}
%\end{equation}
%The quantity leaked at $ t = \SI{240}{\second}$ corresponds to \SI{14}{\milli\gram} of O$_2$.
%Oxidation rates can be tested by repeatedly weighing boxes in air before transferring them to to the glovebox. 
%Both estimates are $<1\%$ of the smallest $\Delta m$ in Table \ref{tab:table1}.

A study on sorption of atmospheric gases by bulk Li metal\cite{hart_sorption_2016} measured mass gain rates in room temperature, artificial dry air of roughly \SI{0.2}{\milli\gram\per\centi\meter\squared\per\hour}.

The boxes have an internal surface area of \SI{150}{\centi\meter\squared}.
If \SI{6}{\minute} elapses during transfer from the vacuum chamber to the glovebox and the gas is as dry as the artificial dry air (estimated at 3.3\% relative humidity\cite{hart_sorption_2016}), then the mass gain per box is estimated at \SI{3}{\milli\gram}.
This is $<1\%$ of the predicted Li efflux from the third box (\autoref{tab:table1}).

Oxidation rates can be tested by repeatedly weighing boxes in air before transferring them to to the glovebox. 

\subsection{Estimation of heating and cooling times}
For ease of modeling, most of the Li should flow during the temperature soak phase rather than during heating and cooling.
A simple model of box 1, heater 1, and the outer shield was created.
Thermal masses and emissivities (see \autoref{tab:table1}) are estimated.
The heater is run at full power.
Evaporative cooling from the Li flow is according to the choked flow model\cite{goldston_lithium_2016}.
The model is integrated from initial temperatures of \SI{300}{K} until box 1 reaches its operating temperature, \SI{920}{K} (\autoref{tab:table1}).

In cooling, each component starts at its operating temperature.
The He gas feed cools the outer box wall and inner heater wall by convection\cite{incropera_fundamentals_1985}.
A flow of 80 standard \si{L\per\minute} is chosen so the He temperature rise is less than half the heater temperature. 
Convective cooling power is de-rated to 30\% to make a conservative estimate for cooling time.

\begin{figure}
\includegraphics[width=\columnwidth]{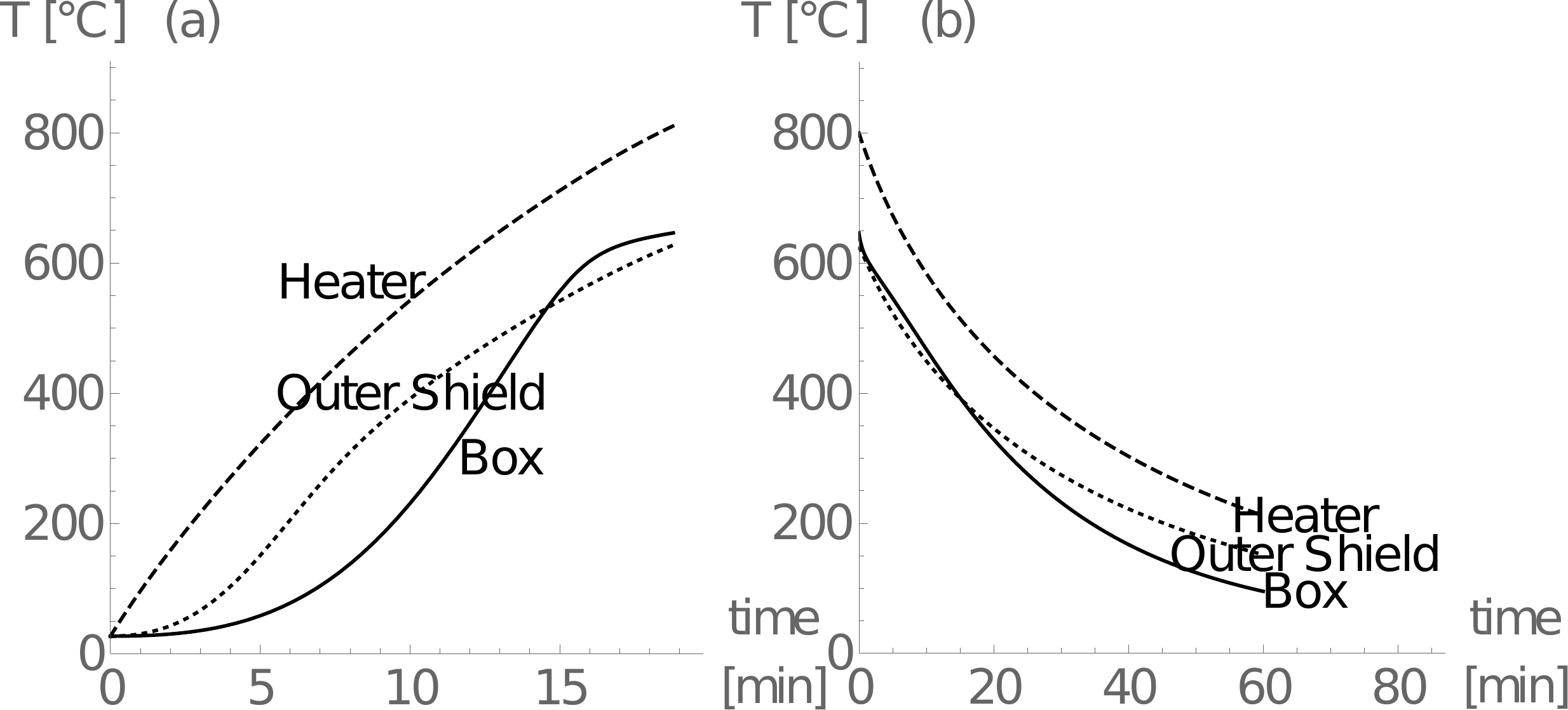}% Here is how to import EPS art
\caption{\label{fig:heatcool} Simple models of box 1 (a) heating and (b) cooling.
The heating model includes radiation transfer from the heaters to box and outer shield, and losses due to latent heat of evaporation from Li flowing out of box 1. The cooling model additionally includes convective cooling from a gas feed.}
\end{figure}

\autoref{fig:heatcool} shows predicted temperatures during heating and cooling.
Integrated Li flow during heating and cooling is 13\% of that during a 30 minute temperature soak.
The transient flow could be measured by an experiment with no soak time.

\section{\label{sec:discussion}Discussion}

The method of inferring mass flow by measuring mass changes of demountable boxes is motivated by the reactivity of Li with many materials.

One disadvantage is that measurements are global (integrated over wall location and time) rather than local.

Local measurements of density or flow rate by optical methods are not trivial as Li could deposit on nearby cool optics, though heated windows\cite{slabinski_lithium_1971}, windows with helium cover gas\cite{vidal_heatpipe_1969}, or a system where the optical paths travel through high-aspect ratio tubes with heated walls and/or cold traps could be possible. Note that the experiments in Reference \onlinecite{slabinski_lithium_1971} were performed at lower temperature.

Measurements inside the boxes involving Li deposition on a small surface over time (e.g.\ a witness plate) can be envisioned but may require detailed knowledge of the surface temperature.
Solidified Li could also prevent easy removal from a mounting surface, and protruding devices would perturb the axisymmetric geometry.

Real-time fluxes could be meausured outside box 3 by measuring the electrical resistance of a thin layer of deposited liquid Li on a plate.
A quartz crystal microbalance could measure vapor flux outside box 3, but vapor fluxes as well as temperatures \textit{inside} the box are likely too high, and Li compatibility would be difficult.

Accurate measurements of box wall and liquid Li surface temperatures are required. If Type N TCs cannot meet accuracy requirements due to drifts, noble metal TCs are a fallback position. Resistance temperature detectors could be used, though they may cover larger surface areas and the number of wires is increased.
Near-infrared measurements of known-emissivity surfaces through holes drilled in the heaters are possible, but would require detailed modeling of radiation (including reflections) inside the device.

\vspace{-10pt}
\section{\label{sec:conclusions}Conclusions}

The design for an experiment to test Li evaporation, flow, and condensation in a vapor box geometry and to validate DSMC models has been presented.

Net flow of Li vapor into each box is measured by disassembling the string of boxes and weighing them separately in an argon glovebox.
The analytical balance used is capable of weighing the \SI{165}{g} boxes to a few \si{\milli\gram}, less than 1\% of the expected efflux from box 3 of \SI{700}{\milli\gram}.

Thermocouples outside boxes are used to infer temperatures inside box walls.
Estimated errors sum to \SI{4.6}{\kelvin}, larger than the $\pm\SI{4.5}{\kelvin}$ corresponding to 10\% changes in Li evaporation flux at \SI{920}{\kelvin}.
Additional work including thermal modeling and estimates of correlations between terms will be necessary to constrain errors for temperature measurements.

\vspace{-10pt}
\begin{acknowledgments}
We acknowledge the help of expert technicians, especially A. Carpe, S. Jurczynski, G. Smalley, J. Taylor, and R. Yager, students including R. Cohen, and valuable discussions with D. Cai, T. Gray, R. Kaita, and R. Majeski.
This work is supported by U.S. Department of Energy Contract No. DE-AC02-09CH11466.
See \url{http://arks.princeton.edu/ark:/88435/dsp01bz60cz987} for the digital data in this paper.
\end{acknowledgments}

\appendix

%\nocite{*}
\bibliography{zot20180803}% Produces the bibliography via BibTeX.

\end{document}